# Revisiting the Sulfur-Water Chemical System in the Middle Atmosphere of Venus

**Wencheng D. Shao[1], Xi Zhang[1], Carver J. Bierson[1] and Therese Encrenaz[2]**

[1] Department of Earth and Planetary Sciences, University of California, Santa Cruz, CA 95064, USA.

[2]LESIA, Observatoire de Paris, PSL University, CNRS, Sorbonne University, University Sorbonne Paris City, 92195 Meudon, France

Corresponding author: Wencheng D. Shao (wshao7@ucsc.edu)

**Key Points:**

- We found that there is no bifurcation behavior in the sulfur-water chemical system as previously claimed.

- The observed SO2-H2O anti-correlation can be explained by the sulfur-water chemistry with mixing ratio variations at the middle cloud top.

- This suggests that the observed temporal variations of SO2 and H2O are linked to the lower atmospheric processes.

## Abstract

Sulfur-water chemistry plays an important role in the middle atmosphere of Venus. Ground based observations have found that simultaneously observed $SO_2$ and $H_2O$ at ~64 km vary with time and are temporally anti-correlated. To understand these observations, we explore the sulfur-water chemical system using a one-dimensional chemistry-diffusion model. We find that $SO_2$ and $H_2O$ mixing ratios above the clouds are highly dependent on mixing ratios of the two species at the middle cloud top (58 km). The behavior of sulfur-water chemical system can be classified into three regimes but there is no abrupt transition among these regimes. In particular, there is no bifurcation behavior as previously claimed. We also find that the $SO_2$ self-shielding effect causes $H_2O$ above the clouds to respond to the middle cloud top in a non-monotonic fashion. Through comparison with observations, we find that mixing ratio variations at the middle cloud top can explain the observed variability of $SO_2$ and $H_2O$. The sulfur-water chemistry in the middle atmosphere is responsible for the $H_2O$-$SO_2$ anti-correlation at 64 km. Eddy transport change alone cannot explain the variations of both species. These results imply that variations of species abundance in the middle atmosphere are significantly influenced by the lower atmospheric processes. Continued ground-based measurements of the co-evolution of $SO_2$ and $H_2O$ above the clouds and new spacecraft missions will be crucial for uncover the complicated processes underlying the interaction among the lower atmosphere, the clouds and the middle atmosphere of Venus.

## Plain Language Summary

Sulfur chemistry composes one important chemical cycle in the Venusian atmosphere. Sulfur dioxide, the most abundant sulfur species, is transported from the lower atmosphere (below the clouds) to the middle atmosphere. On the dayside, sulfur dioxide is dissociated by ultra-violet light and forms various sulfur-bearing species. These species, like polysulfur and sulfuric acid, are critical for the formation of Venus' haze and the sulfuric-acid clouds. Sulfur dioxide is observed to vary by orders of magnitude, though mechanisms underlying those variations remain elusive. In this work, we use a one-dimensional photochemical model to explain the co-evolution of sulfur dioxide and water from ground-based observations. We find that sulfur chemistry and variations inside the clouds are two important factors affecting temporal variations of sulfur dioxide and water. Our study highlights the importance of the interaction among the lower atmosphere, the clouds and the middle atmosphere on Venus.

## 1 Introduction

The sulfur cycle is one major part of the complicated chemistry in the Venus atmosphere. (Yung & DeMore, 1982). Sulfur oxides react with water and form the sulfuric acid clouds at 60-70 km (Young, 1973; Hansen & Hovenier, 1974). Those clouds block the ultraviolet photons globally and separate the Venus atmosphere into two distinct regions in terms of tracer transport and chemistry. The lower atmosphere is characterized by thermochemistry and vigorously convective mixing. The upper part, usually termed as the middle atmosphere ranging from 60 to 100 km, is stably stratified, and photochemistry plays an important role. Previous work (e.g., Yung & DeMore, 1982; Mills, 1998; Zhang et al., 2012; Krasnopolsky, 2012, 2013, 2018) used one-dimensional photochemistry-transport models to explain species abundances in the middle atmosphere. Those models can explain vertical profiles of species like HCL, OCS, $SO_2$ and SO. But not much effort has been put forth on explaining variability of those species.

The Pioneer Venus spacecraft and the International Ultraviolet Explorer observed that $SO_2$ mixing ratio at the cloud top (~70 km) decreases by an order of magnitude during 1970s and 1980s (Esposito, 1984; Esposito et al., 1988; Na et al., 1990). The SPICAV instrument onboard the Venus Express spacecraft observed a secular increase in $SO_2$ at the cloud top between mid-2006 and 2007 (Marcq et al., 2013) and then an overall decrease from 2007 to 2014 (Marcq et al., 2013, 2019b; Vandaele et al., 2017). Ground-based observations from the TEXES high-resolution imaging spectrometer at the NASA Infrared Telescope Facility (IRTF) also detected long-term variations of $SO_2$ at 64 km (near the cloud top) in 2012-2019 (Encrenaz et al., 2016, 2019a). The SPICAV and the TEXES data also show that $SO_2$ above or near the cloud top has large short-term and spatial variations (Encrenaz et al., 2012, 2013, 2016; Vandaele et al., 2017). For water, ground-based telescopes found temporal variations in the disk-integrated $H_2O$ abundance (Sandor & Clancy, 2005; Encrenaz et al., 2016, 2019a). But TEXES discovered that $H_2O$ at 64 km, unlike $SO_2$, exhibits relatively uniform spatial distribution over the Venus disk (Encrenaz et al., 2012, 2013).

Although one-dimensional models can explain the observed vertical profiles of $SO_2$ in the middle atmosphere through eddy diffusion and photochemistry (e.g., Yung and DeMore 1982;

Mills 1998; Zhang et al., 2012; Krasnopolsky, 2012, 2013, 2018), mechanisms underlying horizontal and temporal variations of sulfur species and water are still not well understood. Proposed explanations include middle-atmospheric photochemistry (e.g., Parkinson et al. 2015; Vandaele et al., 2017) and flux variations from the lower atmosphere due to either periodic volcanic injections (e.g., Esposito, 1984; Esposito et al., 1988) or atmospheric dynamical fluctuations (e.g., Cottini et al., 2012; Marcq et al., 2013). Discriminating these mechanisms requires detailed sulfur-water chemical models and detailed observations in high temporal and spatial resolutions.

Encrenaz et al. (2019b; 2020) simultaneously observed variations of $SO_2$ and $H_2O$ at 64 km. These observations range from 2012 to 2019 and are made by TEXES in the spectral range around 7.4 $\mu m$. These observations show not only temporal variations of disk-integrated abundances but also a seemingly temporal anti-correlation between $SO_2$ and $H_2O$. The evidence of this anti-correlation is not very clear in Encrenaz et al. (2019a), but with more data taken recently the correlation is stronger (Encrenaz et al., 2019b, 2020). The cause for this anti-correlation is unknown. Parkinson et al. (2015) (hereafter P15) used the one-dimensional chemistry-diffusion model in Zhang et al. (2012) to study the sulfur-water chemical system in the middle atmosphere. It was found that the system is extremely sensitive to the middle cloud top mixing ratios of $SO_2$ and $H_2O$ at 58 km. But mechanisms of this sensitivity are not well explored. Bierson and Zhang (2020) used a photochemical model describing the full atmosphere of Venus and pointed out that sulfur species abundances in the middle atmosphere are very sensitive to the vertical transport in the lower and middle clouds. The new TEXES data provide a unique opportunity to revisit the sulfur-water chemical system and understand the co-evolution of $SO_2$ and $H_2O$ in the Venus atmosphere in detail.

Using a one-dimensional chemistry-diffusion model, we explore the mechanisms underlying the anti-correlation and variations of $SO_2$ and $H_2O$ from TEXES in this study. We find that the sulfur-water chemical system has three chemical regimes. We show that there is no chemical bifurcation claimed in previous studies (e.g., P15). We also point out that the $SO_2$ self-shielding effect plays an important role in this system. Combining our model with the TEXES data, we find that sulfur chemistry in the middle atmosphere accounts for the long-term anti-correlation of $SO_2$ and $H_2O$. Eddy mixing variations alone cannot produce the observed anti-correlation of both species. The temporal variations of $SO_2$ and $H_2O$ at the observed altitude (64 km) are linked to variations of mixing ratios and fluxes at the middle cloud top (58 km). This implies the observed variability probably originates from processes inside the clouds or from the lower atmosphere.

This paper is organized as follows. First we will introduce our model in section 2. In section 3 we show results from simulations and demarcate different chemical regimes, followed by explanations of mechanisms. In section 4, we compare our simulations with the TEXES data and explore possible mechanisms that could cause the observed variations and anti-correlation.

## 2 Model Description

In this study we use the JPL/Caltech Kinetics Venus model (e.g., Yung & DeMore, 1982; Mills, 1998; Zhang et al., 2012), as was used in P15. This chemical kinetics model has 51 species, 41 photodissociation reactions and over 300 neutral reactions. The model details are described in Zhang et al. (2012). This is the same model used by P15, and following P15 we set the lower boundary at 58 km, the middle cloud top (Knollenberg & Hunten, 1980). Below this altitude the middle and lower cloud layers have low static stability, while above it the Venus atmosphere is stably stratified (Tellman et al., 2009; Imamura et al., 2017; Limaye et al., 2018). To explore the parameter space of the sulfur-water chemical system, we vary lower boundary mixing ratios of two parent species—$SO_2$ and $H_2O$—that are transported upward from the middle cloud region. The range of $SO_2$ lower boundary mixing ratio is 1-75 ppm and that of $H_2O$ is 1-35 ppm, covering the ranges of two species in P15. The temperature, pressure, total number density and eddy diffusion profiles are all the same as Zhang et al. (2012) and P15.

Above 80 km, the volume mixing ratio of $SO_2$ has been observed to increase with height, implying a high-altitude sulfur source (e.g., Sandor et al., 2010; Belyaev et al., 2012). This source may be sulfuric acid or poly-sulfur species (Zhang et al., 2010; 2012). In this study we only use sulfuric acid as the upper sulfur source, same as in P15. Note that the amount of sulfuric acid in the upper atmosphere required to match the sulfur inversion does exceed the upper limits from ground based observations (Sandor et al., 2012). But using poly-sulfur instead or even not including any upper sulfur source (thus no inversion above 80 km) does not alter our conclusions in this paper (see discussions in section 4.1).

The chemical model we use in this study solves the one-dimensional atmospheric continuity equation:

$$\frac{\partial X}{\partial t} = e^{\xi}\frac{\partial}{\partial z}\left(e^{-\xi}K_{zz}\frac{\partial X}{\partial z}\right) + \frac{P-L}{n} \quad (1)$$

(e.g., Zhang et al., 2013). Here $X$ is the volume mixing ratio of a chemical species. $\xi = z/H$, and H is the pressure scale height of the background atmosphere. $K_{zz}$ is the eddy diffusivity. $P$ and $L$ are the total production and loss rates respectively. $n$ is the number density of the background atmosphere. The first and second terms in the right hand side are the eddy diffusion and net production respectively. The molecular diffusion is ignored below the homopause, ~125 km, on Venus. When the chemical system reaches the steady state, the eddy diffusion and net production should balance each other for every species.

The chemistry-diffusion system is usually numerically stiff as the chemical reaction rates could differ by several orders of magnitude. This system can be solved using an implicit Euler time stepping scheme, allowing the time step to exponentially increase in the time marching. A traditional convergence criterion is to check the abundance differences of the chemical species in two successive time steps. If the relative difference is sufficiently small, the model is considered to have reached the steady state. This was the criterion used in P15. Using the same model setup and the traditional convergence criterion, we can reproduce the simulation results of P15 (Fig. S2 (a-c)). However, we found that some cases do not actually reach the steady states defined by Equation (1), i.e., the eddy diffusion and net production terms do not exactly balance each other. See the supplementary materials Text S1 and Fig. S1 for an example case and discussions.

To ensure that the model simulations reach the real steady state, in this study we reinforce a more rigorous convergence criterion, i.e., eddy diffusion and net production must equal in Equation (1) for every species. See an example case in the supplementary materials. Under this new criterion, we reached different simulation results and conclusions from that of P15 (see Fig. S2), as discussed in the following section.

### 3 SO$_2$ and H$_2$O Variability above the Clouds

Following P15, Fig. 1 shows SO$_2$, H$_2$O and SO$_3$ mixing ratio variations (hereafter "maps") at 80 km as a function of SO$_2$ and H$_2$O mixing ratios at 58 km (the middle cloud top).

Here we adopt the same parameter setting as that in P15 to make comparison. In P15's Fig. 9 and 10, $SO_2$ and $H_2O$ maps are "anti-symmetric" across mixing ratio ranges at 58 km (also see Fig. S2 (a-b) in this study). P15 found two regimes, high-$SO_2$-low-$H_2O$ and low-$SO_2$-high-$H_2O$. The transition between the two regimes is abrupt and is called the "chemical bifurcation". However, in our work the $SO_2$ and $H_2O$ maps (Fig. 1; Fig. S2 (d-f)) do not have these behaviors. First, the $SO_2$ and $H_2O$ maps are not "anti-symmetric". Second, it appears no chemical bifurcation or abrupt transition. Instead, the most salient feature is the non-monotonic behavior of $H_2O$ variations as a function of $SO_2$ at 58 km (Fig. 1b).

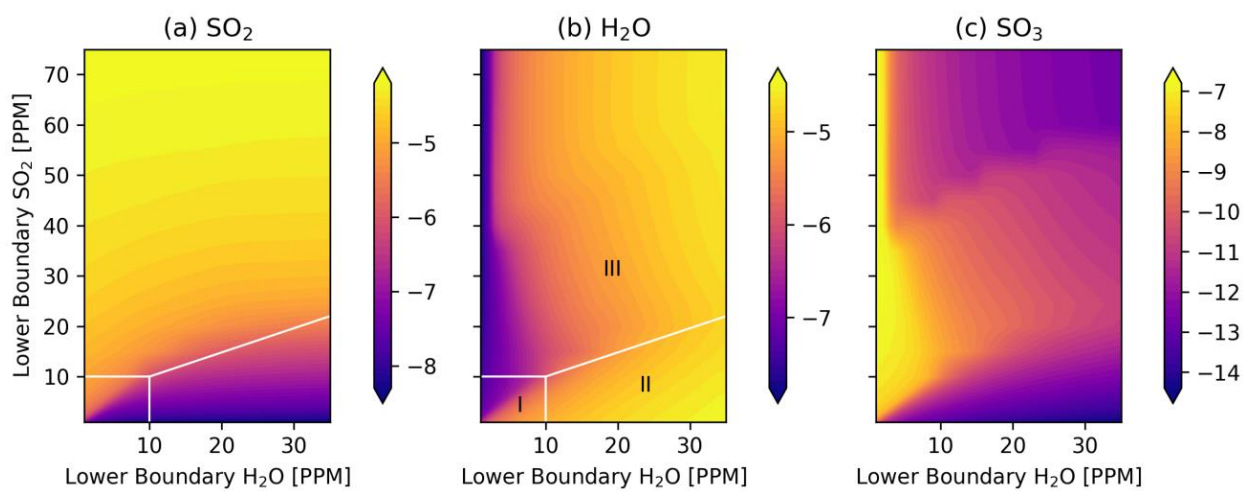

**Fig. 1**. Mixing ratios of $SO_2$ (a), $H_2O$ (b), and $SO_3$ (c) at 80 km as functions of $SO_2$ and $H_2O$ mixing ratios at 58 km. The white lines divide species maps into three regimes, I, II and III. Colors are volume mixing ratios on a logarithmic scale.

According to Fig. 1, we summarize behaviors of the chemical system at 80 km within three regimes:

I.   Low-$SO_2$-low-$H_2O$. $H_2O$ mixing ratio decreases as $SO_2$ at 58 km increases. $SO_2$ mixing ratio decreases as $H_2O$ at 58 km increases. This pattern is similar to that of P15 except that two species do not have very abrupt changes in this regime, i.e., no "chemical bifurcation".

II.  Low-$SO_2$-high-$H_2O$. $H_2O$ is oversupplied. $H_2O$ mixing ratio still decreases as $SO_2$ at 58 km increases. $SO_2$ mixing ratio remains relatively low and insensitive to changes in $H_2O$ at 58 km. $H_2O$ behavior is similar to that in regime I, but $SO_2$ behavior is different.

III.  High-$SO_2$. $H_2O$ mixing ratio increases as $SO_2$ at 58 km increases. $SO_2$ mixing ratio decreases as $H_2O$ at 58 km increases. $SO_2$ behavior is similar to that regime I, but $H_2O$ behavior is different from that in regime I.

In all three regimes, the mixing ratio of an individual species at 80 km increases as its own mixing ratio at 58 km increases. This is primarily a result of eddy diffusive transport from the lower boundary. We also found that $SO_3$ and $H_2O$ maps (Fig. 1b and 1c) exhibit an anti-correlated pattern. This is due to sulfuric acid formation:

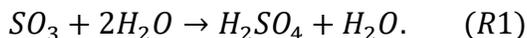

$$SO_3 + 2H_2O \rightarrow H_2SO_4 + H_2O. \qquad (R1)$$

This reaction is the main chemical sink for both $SO_3$ and $H_2O$ near and insides the clouds. This reaction says that two species consume each other, and increase of one species can cause decrease of the other.

Fig. 1b shows that $H_2O$ responds non-monotonically to changes in $SO_2$ at 58 km. From regime I or II to regime III, $H_2O$ at 80 km first decreases and then increases as $SO_2$ at 58 km increases. This behavior is due to the $SO_2$ self-shielding effect. $SO_3$ and $H_2O$ consume each other via reaction (R1). In all regimes, when $SO_3$ above the clouds decreases, $H_2O$ above the clouds increases. $SO_3$ above the clouds is mainly produced by $SO_2$ oxidation:

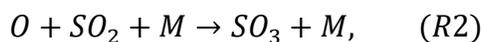

$$O + SO_2 + M \rightarrow SO_3 + M, \qquad (R2)$$

where $M$ is the background atmosphere. Reaction (R2) suggests that $SO_3$ production is affected by abundances of both atomic oxygen O and $SO_2$. The atomic oxygen O above the clouds is mainly produced by $SO_2$ photolysis:

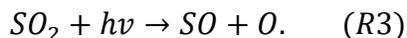

$$SO_2 + hv \rightarrow SO + O. \qquad (R3)$$

Reaction (R3) suggests that atomic oxygen production is affected by $SO_2$ abundance and the amount of photons (or UV light intensity).

When $SO_2$ at the middle cloud top increases, $SO_2$ above the clouds increases. This increase produces more atomic oxygen via $SO_2$ photolysis (R3). $SO_2$ oxidation (R2) then increases $SO_3$ production due to increase of both $SO_2$ and atomic oxygen. Increased $SO_3$ consumes more $H_2O$. This is the chemistry in regimes I and II. Fig. 2 shows an example with $H_2O$ mixing ratio fixed as 10 ppm at 58 km. When $SO_2$ at 58 km increases from 1 to 10 ppm, at 80 km $SO_2$, atomic oxygen and $SO_3$ all increase while $H_2O$ decreases.

As $SO_2$ inside the clouds continues increasing, $SO_2$ at higher altitudes becomes abundant due to eddy transport. Abundant $SO_2$ absorbs many photons, and thus less photons reach lower altitudes such as 80 km. This limits atomic oxygen production via $SO_2$ photolysis (R3). Less atomic oxygen is produced, and this decreases $SO_3$ production via $SO_2$ oxidation (R2). Then sulfuric acid formation (R1) consumes less $H_2O$ and $H_2O$ is accumulated. In this process $SO_2$ absorption and photolysis at higher altitudes "shield" $SO_2$ photolysis at lower altitudes. This is the $SO_2$ self-shielding effect. This effect causes $H_2O$ behaviors in regime III. See the example in Fig. 2 where $H_2O$ at 58 km is fixed as 10 ppm. When $SO_2$ at 58 km increases from 10 to 50 ppm, at 80 km $SO_2$ increases, atomic oxygen and $SO_3$ both decrease due to the $SO_2$ self-shielding, and consequently $H_2O$ increases.

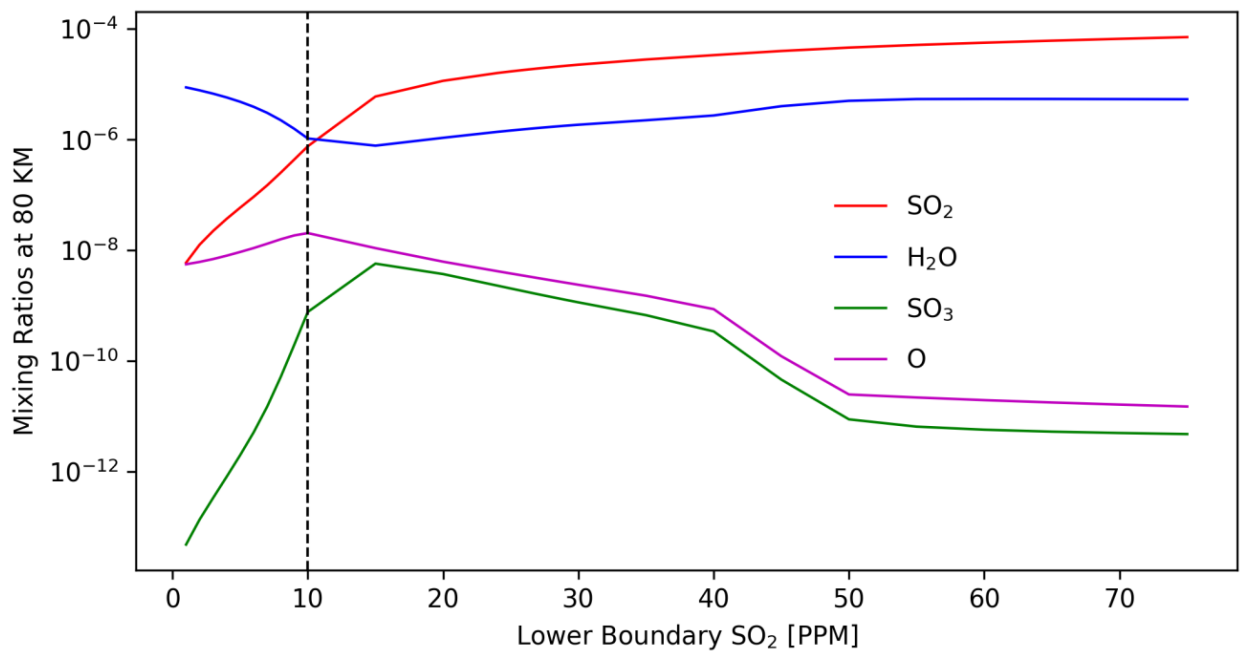

**Fig. 2**. $SO_2$, $H_2O$, $SO_3$ and atomic oxygen mixing ratios at 80 km vary with $SO_2$ mixing ratio at 58 km. $H_2O$ mixing ratio at 58 km is fixed as 10 ppm. The black dashed line delimits regime I and regime III.

SO$_2$ mixing ratio at 80 km generally decreases as H$_2$O at 58 km increases (Fig. 1a). This behavior is strong in regime I, weaker in regime III and almost negligible in regime II. Our analysis suggests that some SO$_3$-involved reactions are key for this behavior. These reactions convert SO$_3$ to SO$_2$ and are here called SO$_3$ pathways:

$$SO_3 + h\nu \rightarrow SO_2 + O, \quad (R4)$$

$$O + SO_3 \rightarrow SO_2 + O_2, \quad (R5)$$

$$S + SO_3 \rightarrow SO_2 + SO, \quad (R6)$$

$$S_2 + SO_3 \rightarrow S_2O + SO_2, \quad (R7)$$

$$SO + SO_3 \rightarrow 2SO_2. \quad (R8)$$

To investigate the role of these SO$_3$ pathways, we do a test in which we shut off all pathways (R4-R8) and see how the chemical system behaves. In our test cases, SO$_2$ lower boundary mixing ratios are all 9 ppm. We designed two groups with (control) and without (experimental) the SO$_3$ pathways, respectively.

Fig. 3 shows mixing ratio variations at 80 km in two groups. In the control group, SO$_2$ at 80 km decreases when H$_2$O at 58 km increases from 1 to 10 ppm. This is because more H$_2$O consumes more SO$_3$ and causes less SO$_2$ produced via the SO$_3$ pathways. When SO$_3$ pathways are not included (the experimental), SO$_2$ mixing ratio is not affected. Consequently SO$_2$ at 80 km remains low and insensitive to variations when H$_2$O at 58 km increases from 1 to 35 ppm.

When H$_2$O mixing ratio at 58 km is above 10 ppm, H$_2$O is oversupplied. In this situation, SO$_3$ mixing ratio is low due to efficient sulfuric acid formation and SO$_3$ pathways (R4-R8) contribute little to the SO$_2$ production. As a result, SO$_2$ at 80 km in both the control and experimental groups remain low and insensitive to H$_2$O variations at 58 km. This explains the SO$_2$ behavior in regime II.

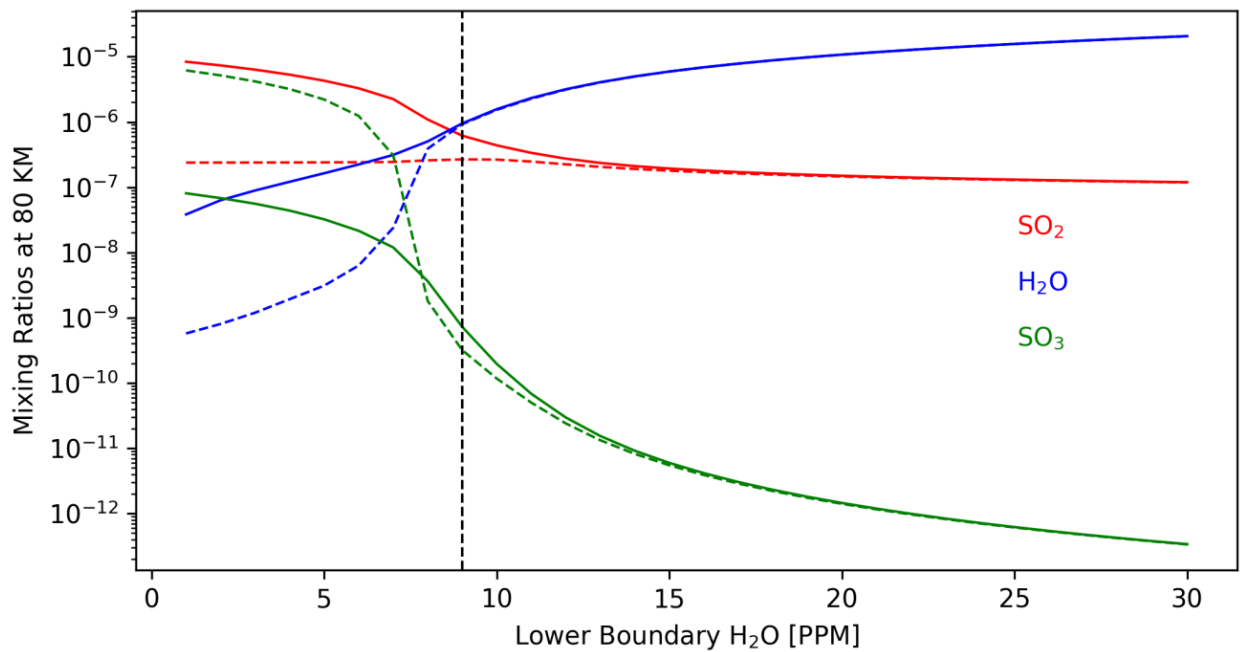

**Fig. 3**. The mixing ratios of $SO_2$ (red), $H_2O$ (blue), and $SO_3$ (green) at 80 km for two groups vary with $H_2O$ mixing ratio at 58 km. $SO_2$ mixing ratio at 58 km is fixed as 9 ppm. In the control group (solid lines) reactions (R4-R8) convert $SO_3$ to $SO_2$, while in the experiment group (dashed lines) these reactions are shut off. The black dashed line delimits regime I and II.

## 4 Comparison with the TEXES data

Encrenaz et al. (2019b; 2020) reported an anti-correlation of disk-averaged abundances of simultaneously observed $SO_2$ and $H_2O$ at ~64 ± 2 km (near the upper cloud top) in 2012-2019 from TEXES/IRTF. As shown in Fig. S5a and Fig. 6a, these data provide unique information about how the two parent chemical species vary together in the middle atmosphere. In particular, the apparent anti-correlation between the disk-averaged $SO_2$ and $H_2O$ could be used to distinguish different proposed mechanisms for the long-term variation. Using our simple one-dimensional chemistry-diffusion model, here we specifically explore two possibilities. First, if the temporal variation of $SO_2$ and $H_2O$ is a result of perturbations below the middle cloud top (e.g. Esposito, 1984; Esposito et al., 1988), varying the lower boundary conditions in our model should be able to reproduce the anti-correlation. On the other hand, if the variation is due to changes in the vertical mixing inside the upper cloud (e.g., Lefevre et al., 2018, 2020), changing the eddy diffusivity in our model should be able to explain the data.

To compare our model with these disk-averaged observations, we set the latitude at $45°N$ and assume that the $SO_2$-$H_2O$ chemical system is in steady state at each individual observational time. The Venus atmosphere is highly variable on timescales of hours and days (e.g., Encrenaz et al., 2013). To isolate longer-period variations (which are more comparable to steady state simulations) we spatially average over the entire disk and temporally average over observations taken within 2 months. The chemical lifetimes of $SO_2$ and $H_2O$ at 64 km are generally less than two months.

### 4.1 Middle Cloud Top Variations

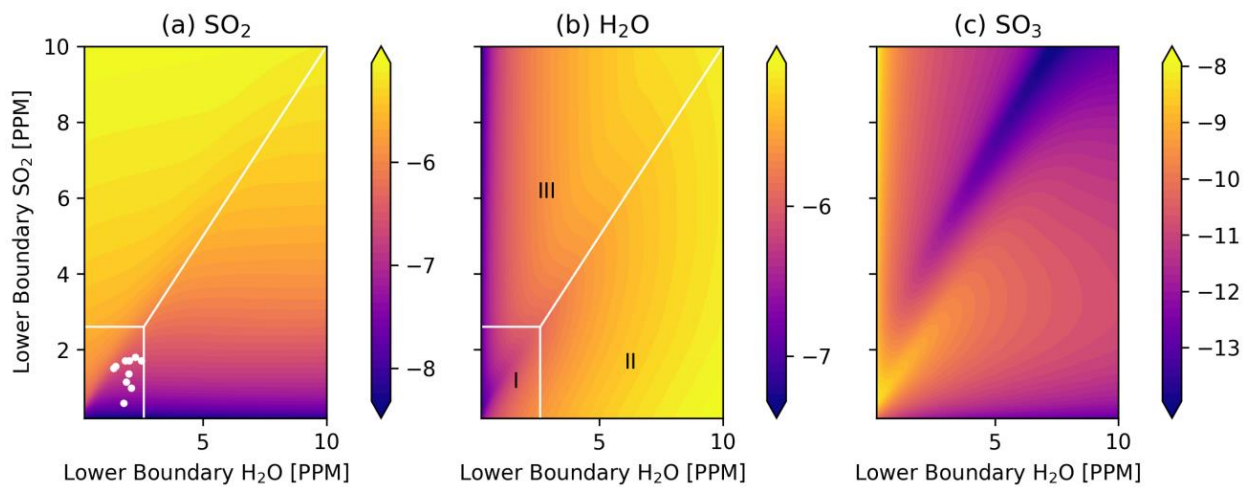

**Fig. 4** Same as Fig. 1, but at 64 km and $45°N$. White dots are models that match the TEXES observations.

To test whether the variations of $SO_2$ and $H_2O$ come from the lower atmospheric processes, we first perform simulations by varying $SO_2$ and $H_2O$ mixing ratios at the middle cloud top (58 km) but fixing the eddy diffusivity. We analyze $SO_2$, $H_2O$ and $SO_3$ variations at 64 km to explore the parameter space (Fig. 4). There are also three chemical regimes at 64 km, similar to that in Fig. 1 where we showed the same species at 80 km. For example, in regime I at both altitudes, $SO_2$ decreases as $H_2O$ at 58 km increases, and $H_2O$ decreases as $SO_2$ at 58 km increases. But the regime boundaries are different between Fig. 1 and 4. The regime I in Fig. 4 only covers $H_2O$ and $SO_2$ at 58 km from 0-2.6 ppm, while regime I in Fig. 1 covers both species at 58 km from 0-10 ppm.

For each individual observational data point of $SO_2$ and $H_2O$ at 64 km, we fix the eddy mixing and carefully tune the $SO_2$ and $H_2O$ mixing ratios at the lower boundary of our model to match the disk-averaged TEXES observations. Fig. 5 shows our simulated $SO_2$ and $H_2O$ mixing ratio profiles. As altitude increases, $SO_2$ decreases by orders of magnitude below 80 km. $H_2O$ remains relatively constant within a factor of 2. These structures are consistent with measurements from SOIR onboard Venus Express (e.g., Belyaev et al., 2012; Bertaux et al., 2007). From 2012 to 2019, $SO_2$ mixing ratio below 80 km varies in a similar way to $SO_2$ at 64 km, consistent with the correlated observations from TEXES and SPICAV (Encrenaz et al., 2019a). $H_2O$ below 80 km also follows the same variation pattern as that at 64 km. In the region between 60 and 80 km, the primary sink for both $SO_2$ and $H_2O$ is sulfuric acid formation (R1). Above 90 km both $H_2O$ and $SO_2$ mixing ratios are supplied by our prescribed sulfur acid source. Note that removing this prescribed source hardly changes $SO_2$ and $H_2O$ profiles below 70 km (Fig. 5). There is some difference at 80 km between cases with and without the source, but this difference does not change three regimes discussed in section 3 except that the transitions among regimes could shift. In this section the mainly focused altitude region is below 70 km, the existence of this source is not important to our results.

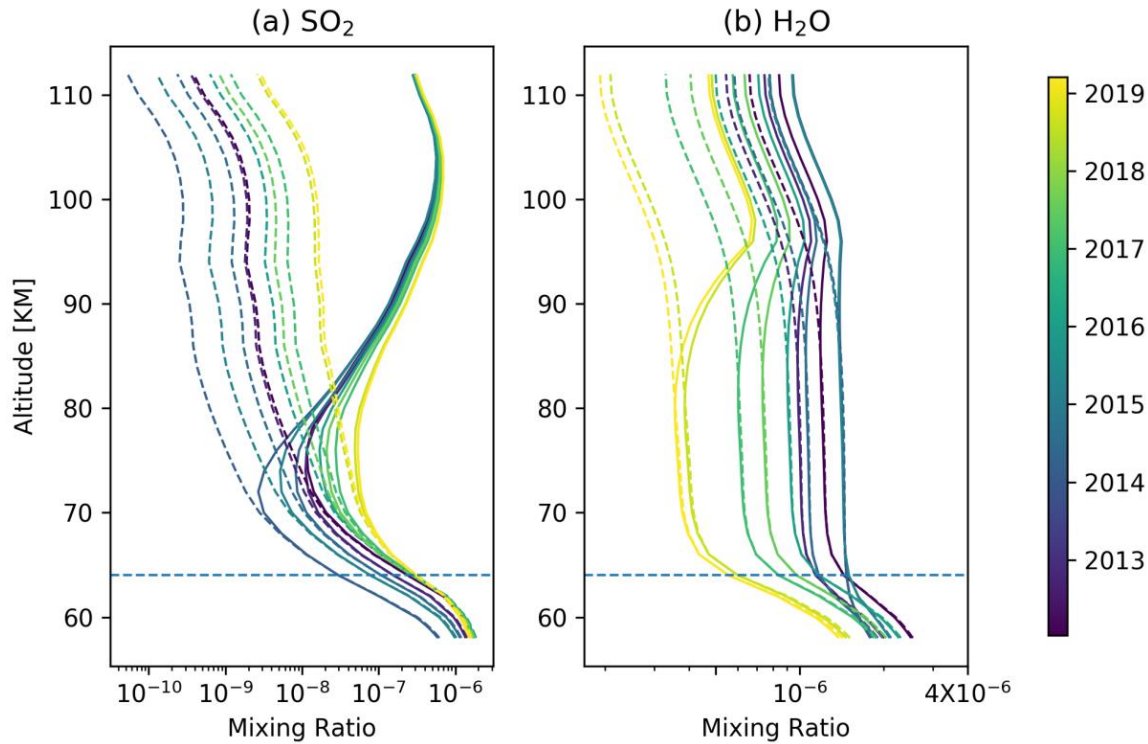

**Fig. 5** Simulated vertical profiles of $SO_2$ (a) and $H_2O$ (b) mixing ratios consistent with the TEXES data at 64 km (horizontal dashed lines). Curves are colored by the observational dates. Solid curves are of cases with the prescribed upper source (sulfuric acid) and dashed curves are of cases without this source (i.e. no sulfuric acid photolysis).

Constrained by the TEXES data, the derived $SO_2$ and $H_2O$ mixing ratios at 58 km are shown in Fig. 6, along with the upward fluxes of two species at 58 km. Mixing ratio variations at 64 km can be divided into two periods: 2012-2015 and 2015-2019 (Fig. 6a; Encrenaz et al., 2019a). In 2012-2015 $SO_2$ at 64 km varies significantly, by a factor up to 5. $H_2O$ varies gently by a factor up to 1.5. In 2015-2019, $H_2O$ varies by a factor over 2, and $SO_2$ remains relatively constant. The two different periods at 64 km also exist at 58 km (Fig. 6b). This similarity between 58 km and 64 km suggests that eddy diffusion plays an important role in the system in addition to photochemistry below 80 km (e.g., Jessup et al., 2015).

The fluxes of two species at 58 km show similar temporal variations to mixing ratios at 64 km (see Fig. S5a and S5c). More interestingly, the two fluxes are strongly and positively correlated (Fig. 6c), and the correlation coefficient is 0.99. This linear-relationship feature for fluxes of two species is a result of the middle-atmosphere photochemistry in our model. Sulfuric acid formation (R1) is the major sink for $SO_2$ and $H_2O$ in the middle atmosphere. If there are no other sinks, then by mass conservation, the two fluxes have to both equal the sulfuric acid formation rate. In fact because some $SO_2$ is also lost to the formation of the poly-sulfur haze, the $SO_2$ flux is larger than the $H_2O$ flux (Fig. 6c).

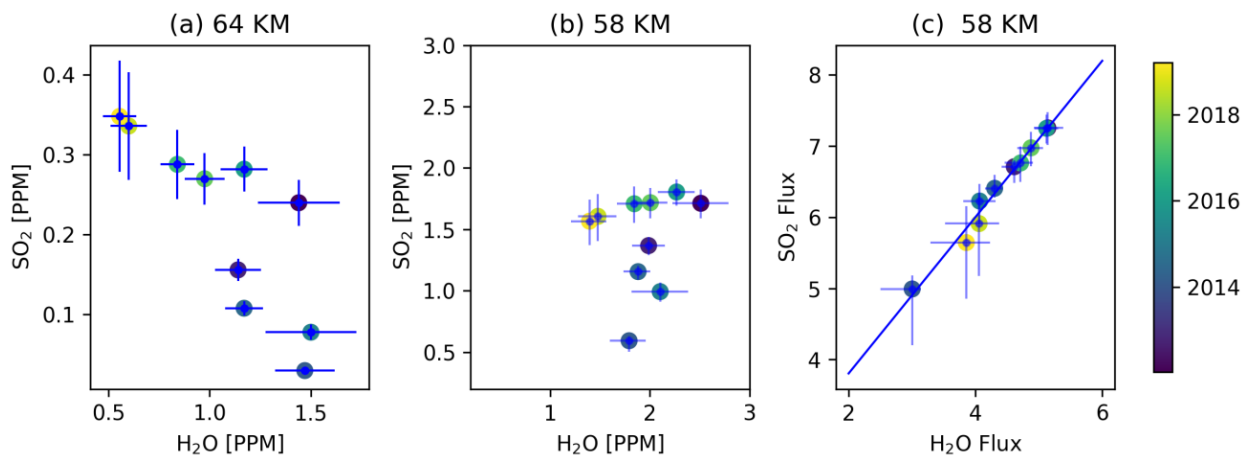

**Fig. 6** Scatter plots of (a) the observed $SO_2$ and $H_2O$ mixing ratios at 64 km, (b) inferred mixing ratios at 58 km and (c) upward fluxes at 58 km. Points are colored by the observational dates. The two fluxes (not considering error bars) can be fit (with 1-sigma error) by $y = (1.10 \pm 0.06)x + (1.61 \pm 0.25)$, where y and x are the $SO_2$ and $H_2O$ fluxes, respectively. Units of x, y and the intercept are $10^{11} cm^{-2} \cdot s^{-1}$. This fit is shown by the solid line in (c). The $SO_2$ and $H_2O$ fluxes are also correlated in this manner in all of our model simulations, not just those shown.

## 4.2 Origin of the Anti-correlation

The TEXES data show that $SO_2$ and $H_2O$ at 64 km are anti-correlated (Fig. 6a). The correlation coefficient is -0.80, and the linear regression slope is -0.27. But the inferred $SO_2$ and $H_2O$ at the middle cloud top (58 km) do not show a good linear correlation (Fig. 6b), implying that the anti-correlation behavior is not universal at all altitudes. This prediction can be tested in future observations. More importantly, the lack of a strong correlation between the two species at 58 km suggests that, although the variations of $SO_2$ and $H_2O$ at 64 km might come from the lower atmospheric processes, the anti-correlation between the two species has a different mechanism.

To diagnose the system, we first notice that the TEXES observations are well located in regime I, i.e., Low-$SO_2$-Low-$H_2O$ (Fig. 4a). In this regime, the $SO_2$ self-shielding effect is insignificant. $SO_2$ and $H_2O$ are linked by $SO_3$ via $SO_2$ oxidation (R2), sulfuric acid formation (R1) and $SO_3$ pathways (R4-R8). More $SO_2$ produces more $SO_3$ that consumes more $H_2O$. More $H_2O$ consumes more $SO_3$ that results in less $SO_2$. It looks that the anti-correlation of $SO_2$ and $H_2O$ might just be a characteristic of the regime I chemistry.

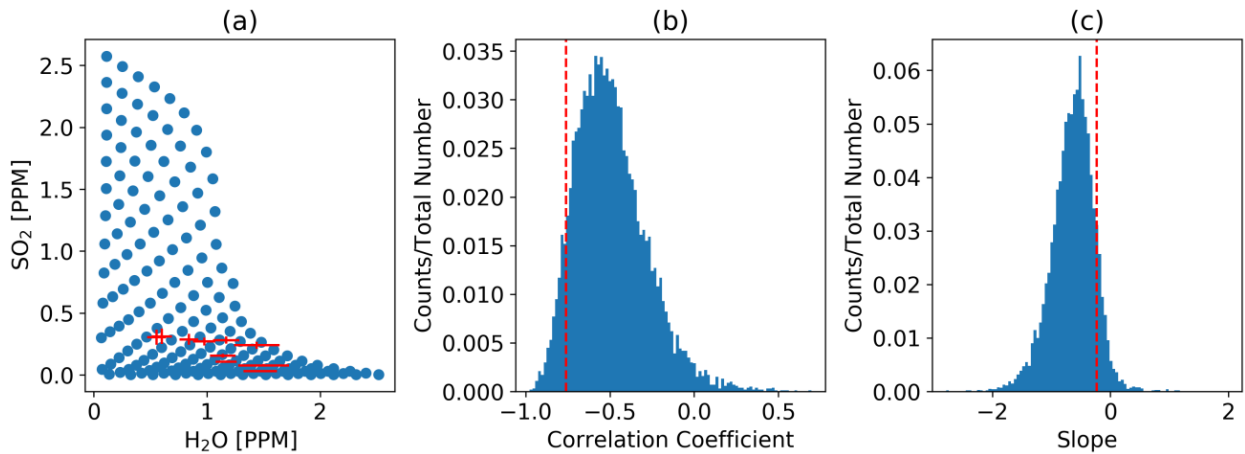

**Fig. 7** (a) Scatter plot of the $SO_2$ and $H_2O$ mixing ratios at 64 km for all cases in the regime I (Fig. 4). Red crosses are the TEXES data with error bars. (b) Statistics of ten-case correlation coefficients and (c) linear regression slopes of $SO_2$ and $H_2O$ at 64 km from our model. In total we have 10 000 correlation coefficients and slopes. We use 100 bins to plot the statistical distributions. The red dashed lines are the values calculated from the TEXES data.

To test this idea, we do a statistical correlation analysis. First we run a suite of model cases in the regime I, and among these cases both $SO_2$ and $H_2O$ at 58 km are evenly spaced by 0.2 ppm between 0.2 and 2.6 ppm. Since we have ten observations, we randomly choose ten cases and calculate the correlation coefficient and the linear slope of $SO_2$ and $H_2O$ at 64 km. We repeat the above analysis for 10 000 times and obtain the distribution of correlation coefficients and slopes (Fig. 7b and 7c). 97.7% of correlation coefficients and slopes are negative, qualitatively consistent with the sensitivity test in Krasnopolsky (2018). The correlation coefficient peaks at -0.5, and the slope peaks at -0.8. This analysis shows that if $H_2O$ and $SO_2$ vary uniformly and independently at the middle cloud top, a negative correlation of $SO_2$ and $H_2O$ at 64 km would be expected due to the regime I chemistry. This suggests that sulfur chemistry in the regime I together with the lower boundary variations can produce the anti-correlation of $SO_2$ and $H_2O$ at 64 km.

Fig. 7 shows that the observed correlation of $SO_2$ and $H_2O$ differs somewhat from the model's prediction. The observed value does not locate at the center of the distributions. Although there are uncertainties in the observations, this discrepancy could also suggest that there are some second-order processes involved. It is likely that the mixing ratios of $SO_2$ and $H_2O$ at 58 km do not follow a uniform distribution as assumed and may be somewhat correlated through atmospheric dynamics inside the middle clouds. Also, the exact location of the center of the distribution might depend on the choice of the eddy mixing profile in the model. But note that changing the eddy mixing alone would not produce the anti-correlation, as detailed below in Section 4.3. Future observations, both remote and in-situ, could help distinguish influences from these factors.

### 4.3 Eddy Mixing Change

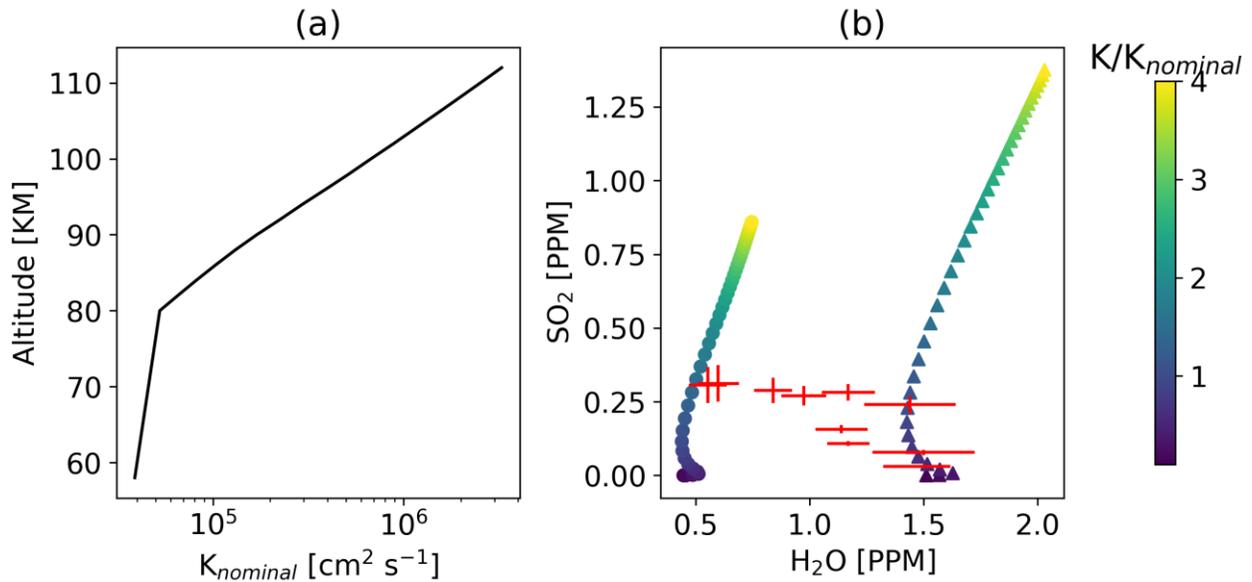

**Fig. 8** (a) The nominal eddy diffusivity profile in our model. (b) Scatter plot of $SO_2$ and $H_2O$ mixing ratios at 64 km for various eddy diffusivity profiles. Diffusivity profiles vary from 0.1 to 4 times the nominal profile, as shown by colors. Circular markers are of cases with 1.0 ppm $SO_2$ and 1.0 ppm $H_2O$ at the lower boundary. Triangular markers are of cases with 1.7 ppm $SO_2$ and 2.5 ppm $H_2O$ at the lower boundary. Red crosses are the TEXES data with error bars.

Now we test whether the middle atmospheric dynamics can produce the variations and anti-correlation of $SO_2$ and $H_2O$ at 64 km. In the above cases we varied the lower boundary conditions and fixed the eddy diffusivity profile (nominal, Fig. 8a). The diffusivity profile is from Zhang et al. (2012), calculated based on measurements from Pioneer Venus (Von Zahn et al., 1979) and radio signal scintillations (Woo & Ishimaru, 1981). Here we fix the lower boundary conditions at 58 km but multiply this diffusivity profile by a constant factor ranging from 0.1 to 4, to explore the influence of the eddy mixing change on species mixing ratio variations at 64 km. We perform two sets of lower boundary conditions. In one set we use 1.0 ppm $SO_2$ and 1.0 ppm $H_2O$ at 58 km. In the other one we use 1.7 ppm $SO_2$ and 2.5 ppm $H_2O$ at 58 km. The results are shown in Fig. 8b. Varying eddy mixing alone above 58 km does change the mixing ratios of $SO_2$ and $H_2O$ at 64 km, but our simulations show that $SO_2$ is far more sensitive to the eddy mixing changes than $H_2O$. This difference in sensitivity of $SO_2$ and $H_2O$ to eddy mixing is qualitatively consistent with Krasnopolsky (2018) and may explain the greater variations of $SO_2$ than $H_2O$ over both space and time observed by TEXES (e.g., Encrenaz et al., 2013) and SPICAV (e.g., Vandaele et al., 2017). When we fix the lower boundary condition and vary the eddy mixing so that $SO_2$ at 64 km changes from 0 to 0.3 ppm, which is approximately the range of the observed $SO_2$ variations, $H_2O$ only varies by less than 30%. In contrast, the observed abundance of $H_2O$ has larger variations up to a factor of ~3. Also, when diffusivity increases in a large range from 0.1 to 4 times the nominal values, both $SO_2$ and $H_2O$ at 64 km generally increase. Thus, we conclude that even though the eddy mixing could vary with time above 58 km, the variations of eddy mixing alone cannot explain the observed variation range of $H_2O$ and the anti-correlation between $SO_2$ and $H_2O$ at 64 km.

## 5 Conclusions and Discussions

In this work we revisited the sulfur-water chemical system in the middle atmosphere of Venus, motivated by the recently and simultaneously observed $SO_2$ and $H_2O$ variations at ~64 km from TEXES/IRTF (Encrenaz et al., 2019b; 2020). Using a one-dimensional chemistry-diffusion model, we studied the co-evolution of $SO_2$ and $H_2O$ in the middle atmosphere of Venus for the first time. We first explored the variability of the chemical species in the system and the underlying mechanisms in a thorough way. We reported chemical regimes and mechanisms different from previous studies. Then we used our model to investigate the long-term anti-correlation of $SO_2$ and $H_2O$ observed by TEXES. We tested two possible mechanisms for the anti-correlation and provided implications of those TEXES observations.

The chemical system is highly dependent on $SO_2$ and $H_2O$ mixing ratios at the middle cloud top at 58 km. $SO_2$ and $H_2O$ mixing ratios above the clouds vary with mixing ratios at 58 km in three regimes: low-$SO_2$-low-$H_2O$ (regime I), low-$SO_2$-high-$H_2O$ (regime II) and high-$SO_2$ (regime III). The pattern of regime I is similar to that in P15 but in a much smaller parameter space. There is no chemical bifurcation or abrupt transition in regime I. In regime II $SO_2$ mixing ratio above the clouds remains low and constant as $H_2O$ at 58 km increases. In regime III, $H_2O$ above the clouds increases as $SO_2$ at 58 km increases, different from $H_2O$ behavior in regimes I or II. Across the regimes there is the non-monotonic variability of $H_2O$ with respect to $SO_2$ variations at 58 km. $H_2O$ and $SO_3$ variations above the clouds are anti-symmetric for all three regimes due to sulfuric acid formation.

The $SO_3$-involved chemistry network connects $SO_2$ mixing ratio above the clouds to $H_2O$ at 58 km. The non-monotonic behavior of $H_2O$ above the clouds results from the interplay among eddy diffusion, neutral chemistry and photolysis processes. In those processes, the $SO_2$ self-shielding effect plays a crucial role.

We explored the mechanisms underlying the variations and anti-correlation between $SO_2$ and $H_2O$ at 64 km from TEXES. We tested two possibilities: eddy mixing change in the middle atmosphere and species variations at the middle cloud top. Both possibilities can originate from lower atmospheric processes. We found that the eddy mixing change alone cannot produce the observed variation range of $H_2O$ or the anti-correlation, while variations of mixing ratios at the middle cloud top with the regime I sulfur chemistry can explain both variations and the long-term anti-correlation of $SO_2$ and $H_2O$. This suggests that the observed $SO_2$ and $H_2O$ variations are more likely due to perturbations on mixing ratios at the middle cloud top rather than changes in the vertical mixing alone.

Although our 1D model provided the first and simplest explanation of the observed anti-correlation behavior of $SO_2$ and $H_2O$, our model is relatively simple with some caveats and could be improved in the future. First, our model does not include phase changes of $H_2O$ and $H_2SO_4$ that could affect trace species abundances in the middle atmosphere (e.g., Gao et al., 2014; Krasnopolsky, 2015). Second, our model does not include the coupling between the lower and middle atmosphere (e.g., Bierson & Zhang, 2020), which can directly link the lower atmospheric processes to the middle atmosphere. Third, our model does not include the local time and spatial variability across latitude and longitude (e.g., Marcq et al., 2019a). It would be important to revisit our proposed mechanism in a more realistic model in the future.

We also noticed discrepancies between different methods and observations of $SO_2$ and $H_2O$ in the middle atmosphere. Several methods have been used to infer the water vapor abundance in the mesosphere of Venus. From Venera 15 data at 30 $\mu m$, Ignatiev et al. (1999) derived a $H_2O$ volume mixing ratio of $10 \pm 2.5$ ppm at $62.5 \pm 1.3$ km for latitudes below 50°. Using ground-based spectroscopy at 3.3 $\mu m$, Krasnopolsky et al. (2013) inferred a $H_2O$ mixing ratio of $3.2 \pm 0.4$ ppm at 74 km for latitudes below 55°. Using VIRTIS abroad Venus Express at 2.6 $\mu m$, Cottini et al. (2015) derived $3 \pm 1$ ppm for $H_2O$ at $69 \pm 1$ km; Fedorova et al. (2016), using SPICAV at 1.38 $\mu m$, inferred $H_2O = 6$ ppm at 62 km; in both cases, low and middle latitudes were observed, and no evidence was found for local time or inter-annual variations. These results are all globally higher than the values inferred by TEXES. A possible reason is the choice of the D/H ratio in the Venus atmosphere. For the TEXES observations, the D/H ratio is taken equal to 200 times the VSMOW (Vienna Standard Ocean Water), following Krasnopolsky (2010), but there is some uncertainty about this parameter. Another reason might be the use of different line transitions and some uncertainty in the line parameters. Finally, in the case of TEXES, the altitude of the penetration level is not precisely defined, since the spectroscopic analysis gives information on the pressure above the continuum level and not the altitude. Nevertheless, it can be seen that all observers agree on the absence of strong local and temporal variations of $H_2O$ at the cloud top of Venus. Indeed, between 2012 and 2016, the $H_2O$ volume mixing ratio inferred by TEXES was more or less constant (Encrenaz et al., 2016), as pointed out by the other teams for the same period.

The main advantage of the method used with TEXES is the fact that both the $SO_2$ and $H_2O$ volume mixing ratios are inferred simultaneously in location and time, at the same penetration level, from the same spectra and the same maps. Thus, the study of their local and temporal variations should not be significantly affected by uncertainty regarding the exact altitude of the penetration level. To our knowledge, TEXES is the only facility that offers this opportunity.

The continued observations of Venus using the TEXES/IRTF instrument will provide us with more information about atmospheric dynamics and tracer transport on Venus. However, due to the limited information in the cloud region and the lower atmosphere, it is still unclear that how dynamics and chemistry in the lower atmosphere, inside the clouds and in the middle atmosphere are coupled and interacted. Besides, the decadal variations of $SO_2$ at the cloud top are observed since 1980s (Esposito, 1984; Esposito et al., 1988; Na et al., 1990; Marcq et al., 2013, 2019b). The underlying mechanism is elusive although our work highlights the importance of the lower atmospheric processes. It would be crucial to continue simultaneously monitoring $SO_2$ and $H_2O$ (as well as other species) in the middle and lower atmosphere, through both ground-based instruments and future spacecraft missions, to provide more clues. A three-dimensional model describing the entire Venus atmosphere involving multiple processes is expected to provide more insights into these problems in the future.

## Data and Acknowledgments

Simulation data from our work are available at https://doi.org/10.6084/m9.figshare.9696476 .

We thank Chris Parkinson, Peter Gao, Yuk Yung and Frank Mills for useful discussions. This work is supported by the China Scholarship Council Fellowship to W. S. and NSF grant AST1740921 to X. Z.. C. J. B. was supported by the National Science Foundation Graduate Research Fellowship under Grant No. NSF DGE 1339067.